\documentclass[11pt]{article}
\usepackage{amsfonts,amssymb,amsmath}

\begin{document}

    \makeatletter
    \@addtoreset{equation}{section}
    \makeatother
    \renewcommand{\theequation}{\thesection.\arabic{equation}}
    \baselineskip 15pt

\title{\bf Nonlocality without inequalities
\footnote{Work supported in part by Istituto Nazionale di Fisica Nucleare,
   Sezione di Trieste, Italy}}
\author{GianCarlo Ghirardi\footnote{e-mail: ghirardi@ts.infn.it}\\
   {\small Department of Theoretical Physics of the University of Trieste,}\\
   {\small Istituto Nazionale di Fisica Nucleare, Sezione di Trieste, and}\\
   {\small International Centre for Theoretical Physics ``Abdus Salam", Trieste, Italy,}\\
   \\ Luca Marinatto\footnote{e-mail: marinatto@ts.infn.it}\\
   {\small Department of Theoretical Physics of the University of
Trieste, and}\\
   {\small Istituto Nazionale di Fisica Nucleare, Sezione di Trieste, Italy,}}

   \date{}

   \maketitle

\begin{abstract}
We prove that every conceivable hidden variable model
 reproducing the quantum mechanical predictions of almost any entangled
 state must necessarily violate Bell's locality condition.
The proof does not involve the consideration of any Bell inequality but it
 rests on simple set theoretic arguments and it works for almost any
 noncompletely factorizable state vector associated to any number of particles
 whose Hilbert spaces have arbitrary dimensionality.
\end{abstract}



\section{Introduction}

It is well known that the natural processes cannot be described
 by resorting to local realistic theories without conflicting with
 the experimentally verified predictions that quantum mechanics attaches
 to measurement outcomes.
The proof of this fact is usually obtained by considering experiments which
 imply a violation of Bell's
 inequalities~\cite{bell1,CHSH}, which are constraints on particular
 combinations of correlation functions that every local theory requires to be satisfied.
Accordingly, experimental evidence of nonlocality is obtained by performing
 different correlation measurements onto sets of particles all prepared
 in identical entangled states~\cite{Aspect}.
In the literature there also exist proofs of nonlocality which do not make use
 of inequalities at all but simply rest on a clever use of EPR-like arguments~\cite{epr}
 and algebraic manipulations of suitable chosen sets of operators.
The first of these ``nonlocality without inequalities" proofs is due to
 D.~M.~Greenberger, M.~Horne and A.~Zeilinger~\cite{ghz}, but the most striking one has
 been exhibited by L.~Hardy~\cite{Hardy} who was able to rule out the existence of any
 local realistic theory accounting for the predictions of any entangled, but not
 maximally entangled, state of two spin-$1/2$ particles.
Subsequent refinements and extensions of Hardy's proof which, due to its simplicity and
 generality, has been defined ``the best version of Bell's theorem''~\cite{merm}, have
 been derived~\cite{gold,hardy1,hardy2}.
The nice aspect of this approach derives from the fact that the experimental
 rejection of any local theory is simply obtained by verifying the occurrence of
 one particular joint measurement outcome~\cite{DeMartiniExp}.
The general theoretical proof of nonlocality we are going to exhibit represents
 a reformulation and a generalization of Hardy's argument valid for almost any
 non-completely factorizable state vector of a system composed of an arbitrary
 number of particles whose Hilbert spaces have arbitrary dimensionality.
The quite general framework we will deal with is that of any conceivable theory
 completing quantum mechanics such that: (i) the most accurate specification it allows of
 the state of the physical system uniquely identifies the probabilities of all single and joint
 measurements outcomes, (ii) such probabilities, when appropriately averaged over, agree
 with the quantum mechanical ones, and (iii) it satisfies the mathematically precise
 Bell locality request~\cite{bell} when the particles are space-like separated.
On the contrary of Hardy's approach, our proof deals with simple set theoretic arguments
 avoiding EPR-counterfactual statements and makes use of the Bell locality condition
 rather than appealing to the Einstein locality condition~\footnote{``But one supposition we should,
 in my opinion, absolutely hold fast: the real factual situation of the system $S_{2}$ is
 independent of what is done with the system $S_{1}$, which is spatially separated from the
 former". A. Einstein in {\em Albert Einstein, Philosopher Scientist}, Ed. by P.A.Schilp, p.85,
 Library of Living Philosophers, Evanston, Illinois (1949).}.
From the experimental point of view, the procedure allows to identify the precise
 measurements which can yield, with a non-zero probability, outcomes putting into
 evidence the unavoidable ``nonlocality" of every conceivable completion of quantum mechanics.
Therefore, contrary to all nonlocality tests based on a violation of some Bell's inequality
 which require different correlation experiments, the tests based on a Hardy-like
 approach require simply the occurrence of a particular joint measurement outcome
 whose probability depends on the entangled state under consideration.
The joint measurement, as we will see by exhibiting two equivalent nonlocality proofs,
 can involve, in general, either complicated multipartite measurements or multiple
 single-particle measurements. In the latter case, the advantage of dealing with
 simple observables is partially reduced by the fact that the probability of
 the desired outcome may be significantly reduced with respect to former case.

This paper is organized as follows: in the second section we make precise
 the notion of a local stochastic hidden variable model while in the third
 section we exhibit the proof of nonlocality without inequalities.
Finally, in the fourth section an equivalent proof is given, having the advantage of
 dealing with single-particle observables.


\section{Local hidden variable models}

Before exhibiting the proof of nonlocality without inequalities, we identify the formal
 framework we will be dealing with throughout the paper and the precise locality request we will use.
The framework consists of the so-called stochastic hidden variable theories for correlation
 experiments of the EPR-Bell type, experiments where observables are measured at spacelike
 separated locations.
The idea underlying such theories is that the most complete specification of the state
 of an individual physical system is represented by the assignment of certain variables $\lambda$,
 which uniquely determine all (single or joint) probability distributions for possible outcomes
 to occur.
More precisely, a stochastic hidden variable model for a correlation experiment, involving
 measurement processes performed on a $n$-partite quantum system described by the state
 $\vert \psi(1,2,\dots,n) \rangle \in {\cal H}_{1}\otimes
 \dots\otimes {\cal H}_{n}$ (where the dimensionality of ${\cal H}_{i}$ is arbitrary), consists of:
 (i) a set $\Lambda$ whose elements $\lambda$ are called hidden variables; (ii) a normalized and
 positive probability distribution $\rho$ defined on $\Lambda$; (iii) a set of probability
 distributions $P_{\lambda}(A_{i}\!=\!a, B_{j}\!=\!b,\dots, Z_{k}\!=\!z)$ for the outcomes
 of single and joint measurements of any conceivable set of observables
 $\left\{A_{i},B_{j},\dots,Z_{k} \right\}$ where each index $\left\{ i,j,\dots,k\right\}$ refers
 to a single particle or to a group of particles, such that:
\begin{equation}
\label{eq0.1} P_{\psi}(A_{i}=a,B_{j}=b,\dots,Z_{k}=z) = \int_{\Lambda}\,d\lambda\,
   \rho(\lambda) P_{\lambda}(A_{i}=a,B_{j}=b,\dots,Z_{k}=z).
\end{equation}
The quantities at the left hand side of Eq.~(\ref{eq0.1}) are the probability distributions
 predicted by quantum mechanics concerning the outcomes $\left\{ a,b, \dots, z\right\}$ for
 the joint measurement of the observables $(A_{i},B_{j}\dots Z_{k})$ respectively, when the
 system is associated the state vector $\vert \psi \rangle$.
A deterministic hidden variable model is a particular instance of a stochastic model
 where all probabilities $P_{\lambda}$ can take only the values $0$ or $1$.

Hidden variable models represent an extremely general class of theories completing quantum
 mechanics by means of the introduction of the parameters $\lambda$, which may be
 completely or only partially accessible, both practically or in principle,
 to the experimenter.
For example the $\lambda$s might simply reduce to the state vector $\vert \psi \rangle$
 of the system, as happens in ordinary quantum mechanics, and in this case no average
 procedure over the distribution $\rho(\lambda)$ is necessary, or they might represent both
 the state vector $\vert \psi \rangle$ and the positions of all the particles of the system,
 as happens in Bohmian mechanics.
In this last case, while $\vert \psi\rangle$ is considered as accessible, the positions
 are assumed to be in principle not accessible so that an average over their distribution
 is needed in order to reproduce the quantum predictions.
 In the most general case, the accessible knowledge about the actual physical situation can
  reduce simply to the knowledge of $\rho(\lambda)$ itself and the physical predictions are
  basically obtained by an averaging procedure.
 An example of such a situation is represented by Bell's model identifying the states of a
  spin-$1/2$ particle by unit vectors in a tridimensional Euclidean space~\cite{bell1}.

Let us now make precise the locality condition which we impose to the theories we are
 interested in.
When consideration is given to measurement processes taking place at
 spacelike separated regions, it is quite natural to require that all
 joint probability distributions satisfy the factorization property
\begin{equation}
\label{eq0.2}
   P_{\lambda}(A_{i}=a,B_{j}=b,\dots,Z_{k}=z) = P_{\lambda}(A_{i}=a)
   P_{\lambda}(B_{j}=b)\dots P_{\lambda}(Z_{k}=z)  \hspace{1cm}\forall
\lambda\in \Lambda\:.
\end{equation}
This factorizability request is commonly referred as Bell's locality condition~\cite{bell},
 as opposed to the Einstein locality.
In the particular case in which the most complete specification of the state of a physical
 system is given simply by the the state vector $\vert \psi \rangle$, i.e., within ordinary
 quantum mechanics, violation of the locality condition of Eq.~(\ref{eq0.2}) may be ascertained
 directly by considering appropriate observables.
In fact, it is well-known that for any entangled state there exist joint probabilities
 which do not factorize.
As a consequence, establishing that ordinary quantum mechanics is a genuine nonlocal theory
 is straightforward, while it is more difficult to prove unavoidable nonlocal structure of any
 conceivable completion of quantum mechanics.
Finally, it is worth noticing that A.~Fine~\cite{fine} has proved that local stochastic hidden
 variable models are completely equivalent to local deterministic hidden variable models.
As a consequence, all single and joint probability distributions $P_{\lambda}$ can be thought to
 assume values within the interval $[0,1]$ or, equivalently, the values $0$ and $1$ only.
Nonetheless, in what follows we will keep on using the general notation $P_{\lambda}$, without
 stating explicitly whether we are dealing with a genuine stochastic hidden variable
 model or not.


\section{Nonlocality without inequalities for almost any entangled state}

Consider an $n$-partite quantum system described by the state vector $\vert \psi(1,\dots,n)\rangle$
  belonging to the Hilbert space ${\cal H}_{1}\otimes \dots \otimes{\cal H}_{n}$, the dimension
  of each ${\cal H}_{i},\:i=1,\dots ,n$ being greater or equal to $2$.
Let us arbitrarily split the set of $n$ particles in two subsets, which we label as $1$ and $2$
 in what follows, and suppose they involve particles located into two space-like separated
 regions.
Finally, we consider the Schmidt decomposition of $\vert \psi\rangle$ in terms of appropriate orthonormal
 sets of states $\left\{ \vert \alpha_{i}(1) \rangle \right\}$ and
 $\left\{ \vert \beta_{i}(2) \rangle \right\}$, belonging to the Hilbert spaces of the first
 and the second group of particles respectively:
\begin{equation}
\label{eq2} \vert \psi(1,2)\rangle = \sum_{i}p_{i}\vert \alpha_{i}(1)\rangle \otimes
   \vert \beta_{i}(2) \rangle\:,
\end{equation}
The weights $p_{i}$ are positive real numbers satisfying the normalization condition
  $\sum_{i}p_{i}^{2}=1$.
Suppose now that the Schmidt decomposition of Eq.(\ref{eq2}) involves at least two different
 weights, which we assume for simplicity to be the first two, so that $p_{1}\neq p_{2}$.
Actually, this is the only hypothesis which is necessary for our proof (in fact, failure of this condition,
 as we will see, will not allow us to conclude anything concerning the existence or not
 of a local stochastic hidden variable model for the state $\vert \psi\rangle$), and it implies
 that we can prove nonlocality only for those non-completely factorizable states whose
 Schmidt coefficients are not all equal.
Consequently, our proof does not apply, for example, to the maximally entangled states of
 $\mathbb{C}^{2}\otimes \mathbb{C}^{2}$ or to the $n$-partite GHZ states.
 To start with, let us define the following two $2\times 2$ unitary matrices $U$ and $V$ whose entries
 depend on the weights $p_{1}$ and $p_{2}$:
\begin{equation}
\label{eq3} U=\frac{1}{\sqrt{p_{1}+p_{2}}}
\begin{bmatrix}
\sqrt{p_{2}} & -i\sqrt{p_{1}} \\
-i \sqrt{p_{1}} & \sqrt{p_{2}}
\end{bmatrix}
\hspace{2cm} V=\frac{1}{\sqrt{p_{1}^{2}+p_{2}^{2}-p_{1}p_{2}}}
\begin{bmatrix}
-i(p_{2}-p_{1}) & \sqrt{p_{1}p_{2}} \\
\sqrt{p_{1}p_{2}} & -i(p_{2}-p_{1})
\end{bmatrix}\:.
\end{equation}
Now, define two orthonormal bases $\left\{ \vert x_{+}(1) \rangle, \vert x_{-}(1)\rangle \right\}$
 and $\left\{ \vert y_{+}(1) \rangle, \vert y_{-}(1)\rangle \right\}$ belonging to the
 two-dimensional linear manifold of the first group of particles spanned by the vectors
 $\left\{ \vert \alpha_{1}(1)\rangle,\vert \alpha_{2}(1)\rangle\right\}$, and
 two bases $\left\{ \vert x_{+}(2) \rangle, \vert x_{-}(2)\rangle \right\}$ and
 $\left\{ \vert y_{+}(2)\rangle, \vert y_{-}(2)\rangle \right\}$ for the two-dimensional
 linear manifold of the second group of particles spanned by the vectors
 $\left\{ \vert \beta_{1}(2)\rangle,\vert \beta_{2}(2)\rangle\right\}$, according to:
\begin{equation}
\label{eq4}
\begin{bmatrix} \vert x_{+}(1) \rangle \\  \vert x_{-}(1) \rangle \end{bmatrix}
=U \begin{bmatrix} \vert \alpha_{1}(1) \rangle \\  \vert \alpha_{2}(1) \rangle
\end{bmatrix} \hspace{1cm}
\begin{bmatrix} \vert y_{+}(1) \rangle \\  \vert y_{-}(1) \rangle \end{bmatrix}
=VU \begin{bmatrix} \vert \alpha_{1}(1) \rangle \\  \vert \alpha_{2}(1)
   \rangle
\end{bmatrix} \hspace{0.1cm}
\end{equation}
\begin{equation}
\label{eq4.01}
\begin{bmatrix} \vert x_{+}(2) \rangle \\  \vert x_{-}(2) \rangle \end{bmatrix}
=U \begin{bmatrix} \vert \beta_{1}(2) \rangle \\  \vert \beta_{2}(2) \rangle
\end{bmatrix} \hspace{1cm}
\begin{bmatrix} \vert y_{+}(2) \rangle \\  \vert y_{-}(2) \rangle \end{bmatrix}
=VU \begin{bmatrix} \vert \beta_{1}(2) \rangle \\  \vert \beta_{2}(2) \rangle
\end{bmatrix} \:.
\end{equation}
The state $\vert \psi \rangle$ of Eq.~(\ref{eq2})can be expressed in three equivalent
 forms by resorting to the basis vectors defined in Eqs.~(\ref{eq4}-\ref{eq4.01}), as:
\begin{eqnarray}
\label{eq5}
   \vert \psi(1,2)\rangle & = & i\sqrt{p_{1}p_{2}}\,[\,\vert x_{+}(1)\rangle
   \vert x_{-}(2) \rangle + \vert x_{-}(1)\rangle  \vert x_{+}(2) \rangle\,]
   +(p_{2}-p_{1}) \vert x_{-}(1)\rangle \vert x_{-}(2) \rangle
   + \sum_{i>2} p_{i}\vert \alpha_{i}(1)\rangle \vert \beta_{i}(2) \rangle
   \nonumber\\
   & = & i\sqrt{p_{1}^{2}+p_{2}^{2}-p_{1}p_{2}}\, \vert y_{-}(1)\rangle
   \vert x_{-}(2) \rangle +i \sqrt{p_{1}p_{2}}\,\vert x_{-}(1)\rangle \vert
   x_{+}(2) \rangle + \sum_{i>2} p_{i}\vert \alpha_{i}(1)\rangle \vert
   \beta_{i}(2) \rangle \nonumber\\
   & = & i\sqrt{p_{1}p_{2}}\,[\,\vert x_{+}(1)\rangle
   \vert x_{-}(2) \rangle + i\sqrt{p_{1}^{2}+p_{2}^{2}
   -p_{1}p_{2}} \vert x_{-}(1)\rangle \vert
   y_{-}(2)  \rangle+ \sum_{i>2} p_{i}\vert \alpha_{i}(1)\rangle \vert
   \beta_{i}(2) \rangle\:.
\end{eqnarray}
With the aim of exhibiting that particular set of joint probability distributions
 which cannot be described by any local hidden variable
 model, we consider the four operator  $X_{1},Y_{1},X_{2}$ and $Y_{2}$.
These are observables having as eigenstates associated to the eigenvalues $+1$ and
 $-1$ the orthonormal vectors $ \left\{ \vert x_{+}(1) \rangle, \vert x_{-}(1)\rangle \right\}$,
 $\left\{ \vert y_{+}(1) \rangle, \vert y_{-}(1)\rangle \right\}$, $\left\{ \vert x_{+}(2) \rangle, \vert
 x_{-}(2)\rangle \right\}$ and $\left\{ \vert y_{+}(2) \rangle, \vert y_{-}(2)\rangle \right\}$
 respectively, while they act as the null operator in the manifolds
 orthogonal to the bidimensional ones corresponding to the non-zero eigenvalues.
According to Eq.~(\ref{eq5}) the quantum joint probabilities concerning the set of
 observables $X_{1},Y_{1},X_{2}$ and $Y_{2}$ satisfy the following relations:
\begin{eqnarray}
 \label{eq7.1}
 P_{\psi}(X_{1}=+1, X_{2}=+1) &=& 0\\
 \label{eq7.2}
 P_{\psi}(Y_{1}=+1, X_{2}=-1) &=& 0 \\
 \label{eq7.3}
 P_{\psi}(X_{1}=-1, Y_{2}=+1) &=& 0 \\
 \label{eq7.31}
 P_{\psi}(Y_{1}=+1, X_{2}=0) &=& 0\\
 \label{eq7.32}
 P_{\psi}(X_{1}=0, Y_{2}=+1) &=& 0 \\
 \label{eq7.4} P_{\psi}(Y_{1}=+1, Y_{2}=+1) &=&
  \frac{p_{1}^{2}p_{2}^{2}(p_{1}-p_{2})^2}{(p_{1}^{2}+p_{2}^{2}-p_{1}p_{2})^{2}} \:.
\end{eqnarray}
Since we have supposed that the (strictly positive) weights $p_{1}$ and $p_{2}$ are such that $p_{1}\neq p_{2}$,
 the joint probability of Eq.~(\ref{eq7.4}) is different from zero and, as we will see,
 this is the crucial relation which will allow us to deny the existence of a local realistic
 description for the state under consideration.
Suppose that a local stochastic hidden variable model reproducing, in accordance with Eq.~(\ref{eq0.1}),
 the quantum predictions for the state $\vert \psi \rangle$, exists.
Accordingly, if we consider for example Eq.~(\ref{eq7.1}), we must have:
\begin{eqnarray}
 \label{eq7.5}
 P_{\psi}(X_{1}=+1, X_{2}=+1) &  = & \int_{\Lambda}d\lambda \rho(\lambda)
 P_{\lambda}(X_{1}=+1, X_{2}=+1) \nonumber \\
 &= & \int_{\Lambda} d\lambda \rho(\lambda) P_{\lambda}(X_{1}=+1)P_{\lambda}(X_{2}=+1)= 0\:,
\end{eqnarray}
where the second equality is implied by the locality condition of Eq.~(\ref{eq0.2}). The last
 equality of Eq.~(\ref{eq7.5}) can be fulfilled if and only if the product
 $P_{\lambda}(X_{1}=+1)P_{\lambda}(X_{2}=+1)$ vanishes almost
 everywhere~\footnote{By the expression ``almost everywhere'' it is meant
 that the argument of the integral may be different from zero at most within a
 non-empty set $\Gamma$ such that $\int_{\Gamma}d\lambda \rho(\lambda)=0$.}
 within $\Lambda$.
An equivalent result holds for Eqs.~(\ref{eq7.2}-\ref{eq7.4}), leading to:
\begin{eqnarray}
 \label{eq8.1}
 P_{\lambda}(X_{1}=+1)P_{\lambda}(X_{2}=+1) &=& 0 \\
 \label{eq8.2}
 P_{\lambda}(Y_{1}=+1)P_{\lambda}(X_{2}=-1) &=& 0 \\
 \label{eq8.3}
 P_{\lambda}(X_{1}=-1)P_{\lambda}(Y_{2}=+1)  &=& 0 \\
 \label{eq8.31}
 P_{\lambda}(Y_{1}=+1)P_{\lambda}(X_{2}=0) &=& 0 \\
 \label{eq8.32}
 P_{\lambda}(X_{1}=0)P_{\lambda}(Y_{2}=+1) &=& 0\\
 \label{eq8.4}
 P_{\lambda}(Y_{1}=+1)P_{\lambda}(Y_{2}=+1) &\neq & 0  \:,
\end{eqnarray}
where the first five equations are supposed to hold almost everywhere within $\Lambda$,
 while the sixth equation has to be satisfied in a subset of $\Lambda$ whose measure
 according to the distribution $\rho(\lambda)$ is non-zero.
Before proceeding with the proof we note that, in the special case where the parameter
 $\lambda$ is the state vector $\vert \psi \rangle$, as happens in ordinary quantum mechanics,
 a violation of the previous constraints is immediately shown.
In fact, given the particular state $\vert \psi \rangle$ of Eq.~(\ref{eq5})
 both $P_{\lambda=\psi}(X_{1}=+1)$ and $P_{\lambda=\psi}(X_{2}=+1)$ are
 different from zero, thus contradicting Eq.~(\ref{eq8.1}).

In order to prove the more general result that no conceivable local stochastic
 hidden variable model can simultaneously satisfy the equations Eqs.~(\ref{eq8.1}-\ref{eq8.4}),
 a manipulations of those equations is requested.
To this end, let us sum Eq.~(\ref{eq8.2}) and~(\ref{eq8.31}) so that, taking into account that
 $P_{\lambda}(X_{2}=-1)+ P_{\lambda}(X_{2}=+0)+P_{\lambda}(X_{2}=+1)=1$, we obtain:
\begin{equation}
 \label{eq8.41}
 P_{\lambda}(Y_{1}=+1)[1-P_{\lambda}(X_{2}=+1)]=0\:.
\end{equation}
Similarly, summing Eqs.~(\ref{eq8.3}) and~(\ref{eq8.32}) we have:
\begin{equation}
 \label{eq8.42}
 [1-P_{\lambda}(X_{1}=+1)]P_{\lambda}(Y_{2}=+1)=0\:.
\end{equation}
Now let us partition the set of hidden variables $\Lambda$ and define the following
 subsets $A$, $B$ and $C$ as:
\begin{eqnarray}
 A &=& \left\{ \lambda\in \Lambda  \vert P_{\lambda}(X_{1}=+1)=0  \right\}, \\
  B & =&\left\{ \lambda\in \Lambda  \vert P_{\lambda}(X_{2}=+1)=0\right\},\\
  C &=& \Lambda-(A\cup B).
\end{eqnarray}
Since $\Lambda-(A\cup B)= (\Lambda-A)\cap (\Lambda-B)$, we have that, for
 all $\lambda$ belonging to $C$, $P_{\lambda}(X_{1}=+1) P_{\lambda}(X_{2}=+1)\neq 0$.
If the set $C$ would have a non-zero measure according to the distribution $\rho$,
 i.e., if $\int_{C}d\lambda \rho(\lambda)\neq 0$, there would be a violation of Eq.~(\ref{eq8.1})
 and, consequently, of Eq.~(\ref{eq7.1}).
Therefore, in order to fulfill Eq.~(\ref{eq8.1}), the set $A\cup B$ must coincide with $\Lambda$
 apart from a set of zero measure, and we are left only with hidden variables
 belonging to either $A$ or $B$.
If $\lambda$ belongs to $A$ then, by definition, $P_{\lambda}(X_{1}=+1)=0$, so that
 Eq.~(\ref{eq8.42}) can be satisfied only if $P_{\lambda}(Y_{2}=+1)=0$.
Equivalently, if $\lambda$ belongs to $B$ then $P_{\lambda}(X_{2}=+1)=0$ and, according to
 Eq.~(\ref{eq8.41}), $P_{\lambda}(Y_{1}=+1)=0$.
Hence, for any $\lambda \in A\cup B$ either $P_{\lambda}(Y_{1}=+1)= 0$ or $P_{\lambda}(Y_{2}=+1)=0$,
 a fact leading to a contradiction of Eq.~(\ref{eq8.4}), which requires that there
 is a set of nonzero $\rho$-measure within $\Lambda$ where both probabilities
 do not vanish.

To summarize, the simple proof we have just presented shows that it is not possible to
 exhibit any stochastic hidden variable model, satisfying Bell's locality condition of
 Eq.~(\ref{eq0.2}), which
 can account for the quantum mechanical predictions of almost any $n$-partite
 quantum entangled state $\vert \psi \rangle \in {\cal H}_{1}\otimes \dots \otimes {\cal H}_{n}$,
 whose Schmidt decomposition, for any splitting of the particles, contains at least
 two different weights.

Exactly like in the original Hardy's proof~\cite{Hardy}, the experimental test of nonlocality for
 the entangled states we are considering, simply consists in testing the occurrence of the joint
 measurement outcomes of Eq.~(\ref{eq7.4}) whose probability
\begin{equation}
 \label{eq8.5}
 P_{\psi}(Y_{1}=+1, Y_{2}=+1) = \frac{p_{1}^{2}p_{2}^{2}(p_{1}-p_{2})^2}{(p_{1}^{2}+p_{2}^{2}-p_{1}p_{2})^{2}}
\end{equation}
does not vanish whenever $p_{1}, p_{2} \neq 0$ and $p_{1}\neq p_{2} $.


\section{An equivalent proof}

To experimentally test the probability of Eq.~(\ref{eq8.5}), one has to perform measurements
 of the multipartite observables $Y_{1}$ and $Y_{2}$.
Since they might involve (possibly large) groups of particles, an experiment like this could be quite
 complicated to perform from a practical point of view.
In order to overcome this problem, we are going to exhibit now a modification of the previous
 proof which makes use of repeated application of the Schmidt decomposition of the $n$-partite
 state in order to identify a non-zero joint probability distribution conflicting with Bell's
 locality, which involves only single-particle observables.
The only drawback of this equivalent proof consists, as we will see, in
 a possibly reduced value for the probability of the outcome conflicting with locality condition,
 which in general becomes smaller when the number $n$ of particles increases.

We start by considering a tripartite system and later we will show how to generalize
 the proof to cover the case of an arbitrary number of particles.
Consider the Schmidt decomposition of a tripartite state $\vert \psi(1,2,3) \rangle$ in
 terms of a set of bipartite orthonormal states of the first and
 the second particle $\left\{ \vert \phi_{k}(1,2)\rangle \right\}$, and of a set of
 orthonormal states of the third particle $\left\{ \vert\tau_{k}(3) \rangle \right\}$:
\begin{equation}
\label{eq10} \vert \psi(1,2,3)\rangle =\sum_{k} q_{k} \vert \phi_{k}(1,2)\rangle
   \otimes \vert \tau_{k}(3)\rangle.
\end{equation}
where $q_{k}\geq 0$ and $\sum_{k}q_{k}^{2}=1$. Suppose now that, within the orthonormal set
 $\left\{\vert \phi_{k}(1,2)\rangle \right\}$, there exists a state, whose associated $q_{k}$
 is different from zero, let us say $\vert \phi_{1}(1,2)\rangle$, such that at least two different
 weights appear in its Schmidt decomposition~\footnote{Actually the
 tripartite states for which our proof holds, are the ones whose
 Schmidt decomposition, performed by grouping the particles in every
 possible manner, contains at least one bipartite entangled
 state for an arbitrary pair of particles which satisfies the above request.}.
By performing the unitary transformations $U$ and $V$ as defined in Eqs.~(\ref{eq4}-\ref{eq4.01}),
 the state $\vert \psi(1,2,3) \rangle$ of Eq.~(\ref{eq10}) can be decomposed in
 the following form:
\begin{eqnarray}
\label{eq11} \vert \psi(1,2,3)\rangle & = & q_{1}\Big( i\sqrt{p_{1}p_{2}}\,[\,
   \vert x_{+}(1)\rangle
   \vert x_{-}(2) \rangle + \vert x_{-}(1)\rangle  \vert x_{+}(2) \rangle\,]
   +(p_{2}-p_{1}) \vert x_{-}(1)\rangle \vert x_{-}(2) \rangle
   \Big) \otimes \vert \tau_{1}(3) \rangle \nonumber \\
& &  + \,q_{1}\Big( \sum_{i>2} p_{i}\vert \alpha_{i}(1)\rangle \vert
   \beta_{i}(2) \rangle \Big) \otimes \vert \tau_{1}(3) \rangle
    + \sum_{k>1} q_{k} \vert \phi_{k}(1,2) \rangle \otimes \vert
   \tau_{k}(3)\rangle\:,
\end{eqnarray}
Moreover, two other decompositions exist in analogy with Eq.~(\ref{eq5}).

Denoting by $T_{3}$ the single-particle observable of the
   Hilbert space of the third particle having the vectors
   $\left\{ \vert \tau_{k}(3) \rangle \right\}$ as its eigenstates associated
   to a set of eigenvalues  $\left\{ t_{k} \right\}$ which we may choose
   so that $t_{1}$ is a non-degenerate eigenvalue, the following quantum
   probability distributions hold for the state of Eq.~(\ref{eq11}):
\begin{eqnarray}
\label{eq12.1}
P_{\psi}(X_{1}=+1, X_{2}=+1, T_{3}=t_{1}) &=& 0\\
\label{eq12.2}
P_{\psi}(Y_{1}=+1, X_{2}=-1, T_{3}=t_{1}) &=& 0 \\
\label{eq12.3}
P_{\psi}(X_{1}=-1, Y_{2}=+1, T_{3}=t_{1}) &=& 0 \\
\label{eq12.31}
P_{\psi}(Y_{1}=+1, X_{2}=0, T_{3}=t_{1}) &=& 0 \\
\label{eq12.32}
P_{\psi}(X_{1}=0, Y_{2}=+1, T_{3}=t_{1}) &=& 0 \\
\label{eq12.4} P_{\psi}(Y_{1}=+1, Y_{2}=+1, T_{3}=t_{1}) &\neq&  0 \:.
\end{eqnarray}
Once again the existence of a local and stochastic hidden variable model reproducing the quantum mechanical
probability distributions
   of Eqs.~(\ref{eq12.1}-\ref{eq12.4}), implies the following relations:
\begin{eqnarray}
\label{eq13.1}
P_{\lambda}(X_{1}=+1)P_{\lambda}(X_{2}=+1)P_{\lambda}(T_{3}=t_{1}) &=& 0 \\
\label{eq13.2}
P_{\lambda}(Y_{1}=+1)P_{\lambda}(X_{2}=-1)P_{\lambda}(T_{3}=t_{1}) &=& 0 \\
\label{eq13.3}
P_{\lambda}(X_{1}=-1)P_{\lambda}(Y_{2}=+1)P_{\lambda}(T_{3}=t_{1}) &=& 0 \\
\label{eq13.31}
P_{\lambda}(Y_{1}=+1)P_{\lambda}(X_{2}=0)P_{\lambda}(T_{3}=t_{1}) &=& 0 \\
\label{eq13.32}
P_{\lambda}(X_{1}=0)P_{\lambda}(Y_{2}=+1)P_{\lambda}(T_{3}=t_{1}) &=& 0 \\
\label{eq13.4}
  P_{\lambda}(Y_{1}=+1)P_{\lambda}(Y_{2}=+1)P_{\lambda}(T_{3}=t_{1})&\neq& 0\:.
\end{eqnarray}
As before, the first five equations must be satisfied almost everywhere
   within $\Lambda$, while the last one has to be satisfied in a set of non-zero
   measure with respect to the distribution $\rho(\lambda)$.

Let us now show that a local stochastic hidden variable model
   satisfying Eqs.~(\ref{eq13.1}-\ref{eq13.4}) cannot exist.
The procedure we are going to follow  consists in splitting the set of hidden variables
   $\Lambda$ into two complementary and disjoint subsets $\Omega_{1}$ and
   $\Omega_{2}$, referring to the possible values of the probability
distribution
   for the outcome $t_{1}$ of the observable $T_{3}$.
They are defined as $\Omega_{1} =\left\{ \lambda\in \Lambda \vert
   P_{\lambda} (T_{3}=t_{1})=0\right\}$ and $\Omega_{2} =\left\{ \lambda\in
   \Lambda \vert  P_{\lambda} (T_{3}=t_{1})\neq 0\right\}$.
Given any value of the hidden variable $\lambda$, two possible cases can
   occur: either $\lambda \in \Omega_{1}$ or $\lambda \in \Omega_{2}$.
If $\lambda \in \Omega_{1}$, the left hand side of Eq.~(\ref{eq13.4}) vanishes
   and the equation cannot be satisfied.
If $\lambda$ belongs to $\Omega_{2}$ where $P_{\lambda}(T_{3}=t_{1})\neq 0$,
   the equations from~(\ref{eq13.1}) to~(\ref{eq13.4}) reduce to
   Eqs~(\ref{eq8.1}-\ref{eq8.4}) respectively.
This being the case, we can apply the previous arguments to conclude that no local
 stochastic hidden variable model exists which can reproduce the quantum probabilities
 for all tripartite states, whose Schmidt decomposition involves at least one bipartite
 state having at least two different weights in its decomposition.
As before, this approach uniquely determines the set of outcomes of those
 joint measurements whose occurrence constitutes the experimental proof of nonlocality.
Contrary to the first method those measurements involve only elementary single-particle
 measurements, which are simpler to perform, but the probability of such
 an event:
\begin{equation}
 P_{\psi}(Y_{1}=+1, Y_{2}=+1, T_{3}=t_{1}) = q_{1}^{2}
 \cdot\frac{p_{1}^{2}p_{2}^{2}(p_{1}-p_{2})^2}{(p_{1}^{2}+p_{2}^{2}-p_{1}p_{2})^{2}}
\end{equation}
is smaller or equal to that of Eq.~(\ref{eq8.5}).

The argument can be generalized in a straightforward way to any number of particles. In fact, given a
four-particle state, we first perform its Schmidt
   decomposition in terms of orthonormal tripartite states $\left\{
   \vert \phi_{k}(1,2,3)\rangle \right\}$ and single-particle states
   $\left\{ \vert \sigma_{k}(4) \rangle \right\}$ and then we check
   whether at least one tripartite state of the decomposition belongs to the
   set of states which we have just proven not to admit a local hidden variable
   description.
If this is the case we define two disjoint and complementary sets of $\Lambda$,
   that is the sets $\Theta_{1}= \left\{ \lambda\in \Lambda \vert
   P_{\lambda} ( S_{4}=s_{1}) =0 \right\}$ and $\Theta_{2}=\left\{ \lambda\in
   \Lambda \vert P_{\lambda}(S_{4}=s_{1})\neq 0\right\}$, where $S_{4}$ is an
   observable of the Hilbert space of the fourth particle whose eigenstates
   are the vectors $\left\{ \vert \sigma_{i}(4)\rangle\right\}$
   and for which $s_{1}$ is not a degenerate eigenvalue.
We then get quantum probability distributions of the form of
   Eqs.~(\ref{eq13.1}-\ref{eq13.4}), where at the left hand side
   the extra multiplicative factor $P_{\lambda}(S_{4}=s_{1})$ appears.
Once again, if $\lambda \in \Theta_{1}$ the analogous expression to Eq.~(\ref{eq13.4})
   containing the extra factor cannot be satisfied, while, when
   $\lambda \in \Theta_{2}$, we get back to the previous case.
This proves that, step by step, we can generalize our argument to any
   number of particles.
Note that the $n$-particle entangled states for which our proof holds are
   those for which, by considering all conceivable Schmidt decompositions
   in terms of bipartite states and $(n-2)$ single-particle states,
   at least one entangled bipartite state involving at least two
different weights in its Schmidt decomposition
   appears.
Such a set contains almost all entangled states of $n$-particles.


\section{Conclusions}
In this paper we have proven that every conceivable (deterministic or stochastic)
 hidden variable model reproducing the quantum mechanical predictions of almost
 any entangled state must necessarily violate Bell's locality condition.
Two proofs of nonlocality have been exhibited, both of them not involving the
 consideration of any Bell inequality but simply resting on straightforward set
 theoretic arguments.



\begin{thebibliography}{99}
\bibitem{bell1} J.S. Bell, {\it Physics} {\bf 1}, 195 (1964).
\bibitem{CHSH} J.F. Clauser, M.A. Horne, A. Shimony, and R.A. Holt,
 {\it Phys. Rev. Lett.} {\bf 23}, 880 (1969).
\bibitem{Aspect} A. Aspect, P. Grangier, and G. Roger, {\it Phys. Rev. Lett.} {\bf 47},
 460 (1981).
\bibitem{epr} A. Einstein, N. Rosen and B. Podolsky, {\it Phys. Rev. }
   {\bf 47}, 777 (1935).
\bibitem {ghz} D.M. Greenberger, M. Horne, and A. Zeilinger in {\it Bell's
   Theorem, Quantum Theory, and Conceptions of the Universe}, M. Kafatos ed.,
   Kluwer, Dordrecht (1989).
\bibitem{Hardy} L. Hardy, {\it Phys. Rev. Lett.} {\bf 71}, 1665 (1993).
\bibitem{merm} N.D. Mermin, {\it Ann. N.Y. Acad. Sci.} {\bf 755}, 616 (1995).
\bibitem{gold} Sheldon Goldstein, {\it Phys. Rev. Lett.} {\bf 72}
   1951 (1994).
\bibitem{hardy1} L. Hardy, in {\it New Developments on Fundamental Problems
   in Quantum Physics}, edited by M. Ferrero and A. van der Merwe, Kluwer,
   Dordrecht, 163 (1997).
\bibitem{hardy2} D. Boschi, S. Branca, F. De Martini, and L. Hardy, {\it
   Phys. Rev. Lett.} {\bf 79}, 2755 (1997).
\bibitem{DeMartiniExp} G. Di Giuseppe, F. De Martini, and D. Boschi,
 {\it Phys. Rev. A} {\bf 56}, 176 (1997).
\bibitem{bell} J.S. Bell, {\it  Bertlemann's socks and the nature of reality}
   in {\it John S. Bell on The Foundations of Quantum Mechanics}, M. Bell,
   K. Gottfried \& M. Veltman eds., World Scientific (2001).
\bibitem{fine} A. Fine, {\it Phys. Rev. Lett.} {\bf 48}, 291 (1982).

   \end{thebibliography}
\end{document}